\documentclass[pra,twocolumn,showpacs,amsmath,amssymb]{revtex4}
\usepackage{graphicx}
\usepackage{dcolumn}
\usepackage{bm}
\usepackage{appendix}
\usepackage{multirow}
\begin{document}

\title{Coupled-cluster calculations of properties of Boron atom as a monovalent system}
\author{H. Gharibnejad}
\affiliation {Department of Physics, University of Nevada, Reno, Nevada 89557, USA.}
\author{A. Derevianko}
\affiliation {Department of Physics, University of Nevada, Reno, Nevada 89557, USA.}

\begin{abstract}
We present relativistic coupled-cluster (CC) calculations of energies, magnetic-dipole hyperfine constants, and electric-dipole transition amplitudes for low-lying states of atomic boron. The trivalent boron atom is computationally treated as a monovalent system.
We explore performance of the CC method at various approximations. Our most complete treatment involves singles, doubles and the leading valence triples.
The calculations are done using several approximations in the coupled-cluster (CC) method.
The results are within 0.2-0.4\% of the energy benchmarks.
The hyperfine constants are reproduced with 1-2\% accuracy.
\end{abstract}

\maketitle

\section{INTRODUCTION}
\label{Sec:Introduction}

Atomic parity violation~\cite{Khr91} (APV) provides powerful constraints on new physics beyond the standard model of elementary particles~\cite{MarRos90,Ram99}. While the experiments are carried out at low energies, the derived constraints are both unique and complementary to those obtained from direct searches for new physics with high-energy particle colliders.
For example, the latest APV results~\cite{PorBelDer09,PorBelDer10} set new mass bounds on exotic new-physics particles, $Z'$  bosons, which are ubiquitous in competing extensions of the standard model. These APV bounds improve upon the earlier results of the Tevatron collider and cut out the lower-energy part of the  discovery reach of the Large Hadron Collider.

 Interpretation of APV experiments in terms of electroweak observables  requires input from atomic-structure calculations. APV is the field where focus on {\em precision} of both atomic experiment and theory is of  importance.
As we look at the entire body of experimental APV results with various atoms, we find that the most accurate measurements were carried out with $^{133}$Cs atoms~\cite{WooBenCho97,BenWie99}. The second best APV measurement was done with $^{205}$Tl~\cite{VetMeeMaj95}.  Ideally, the uncertainty of calculations should match the experimental error bars. For Cs, the 0.27\% uncertainty of relativistic many-body calculations~\cite{PorBelDer09,PorBelDer10} is better than the experimental 0.35\% error bar~\cite{WooBenCho97,BenWie99}. For Tl, however, the
situation is  reversed. The experimental accuracy here is about 1\%~\cite{VetMeeMaj95}, while the theoretical errors are estimated
to 2.5-3\%~\cite{DzuFlaSil87J,HarLinMar90,KozPorJoh01}. While for Cs it took a little over a decade for the theory to match the experimental accuracy, for Tl, even after almost two decades the state of the theory remains inadequate.

The goal of this paper is to start exploring the feasibility of transferring high-precision many-body techniques developed for Cs~\cite{PorBelDer09,PorBelDer10} to the Tl atom. There is a crucial distinction between the Cs and Tl atomic structures. Cs has a single $6s_{1/2}$ valence electron outside the closed-shell core, while Tl nominally has three valence electrons ($6s_{1/2}^2\, 6p_{1/2}$ ground state configuration). Since $6p_{1/2}$ is the only active electron involved in the measured $6p_{1/2}-6p_{3/2}$ PNC amplitude~\cite{VetMeeMaj95}, it is natural to wonder if the $6s_{1/2}^2$ shell could be considered as a part of the core, thereby enabling reuse of the Cs techniques.  The relevant figure of merit is the energy gap between the core and the valence subspaces. At the Dirac-Hartree-Fock (DHF) level, this gap for Cs ($6s_{1/2}-5p_{3/2}$) is $156479\, \mathrm{cm}^{-1}$ while in Tl, when treated as a monovalent system the gap between the $6s_{1/2}$ and the  $6p_{1/2}$ is about 1.5 times smaller.
Considering the smallness of this gap,
the interactions between the outer core-shell and the valence electrons have to be treated in non-perturbative fashion.

Before proceeding to computationally-expensive  81-electron thallium atom, in this paper we focus on a prototype atom, boron.
Boron belongs to the same IIIA group of the periodic table but has only 5 electrons.  The ground state configuration of boron reads
$1s_{1/2}^2 2s_{1/2}^2 2p_{1/2}$, with three valence electrons.
Here, however, we have treated boron as a monovalent atom by placing the two lower energy valence electrons (2$s_{1/2}^2$) in the core. Such a treatment greatly simplifies the underlying equations and calculations. Treating the boron as a monovalent system is also justified by the fact that the $2p$ electron is often the excited electron in optical transitions.

Notice that in the non-relativistic Coulomb approximation, the energies of the $2s$ and $2p$ electrons would be identical, again strongly suggesting that the perturbative monovalent approach would fail right from the onset. The mean-field effects lift this degeneracy resulting in the energy gap of $131292 \, \mathrm{cm}^{-1}$ in the DHF approximation.

Another compelling feature of  boron is that as a  five-electron system it lies at the  applicability border of high-accuracy variational methods~\cite{BubAda11}.
In the case of few-body systems, e.g., He and Li, the computational accuracies have been substantial (see, e.g., Refs.~\cite{NakNak07,YanNorDra08,PucKedPac09}.)  Only very recently \citet{BubAda11} have extended the reach of the full-scale variational methods to boron. While such accurate variational calculations were carried out for boron,
the problem with variational methods is the impracticality of extending them to even larger atoms, such as the 81-electron Tl. That leaves many-body methods as the best hope for accurate computations for Tl.
Due to the strongly-correlated nature of these systems and the desired high accuracy, one has to employ non-perturbative (all-order) methods, where certain classes of diagrams are summed to all orders of many-body perturbation theory in the residual Coulomb interaction between electrons.


In this paper we employ  arguably the  most popular all-order method: the coupled-cluster (CC) method; this method was at the heart of  high-accuracy APV calculations for Cs~\cite{PorBelDer09,PorBelDer10} and other heavy atoms~\cite{ ChaSahDas03}. We will use an {\em ab initio} relativistic formulation.
Qualitatively, various CC methods are distinguished  by the maximum number of simultaneously excited electrons from the reference DHF Slater determinant.  Here we include single and double excitations of core electrons and  single, double, and triple excitations of the core and valence electrons. We refer to this scheme as the CCSDvT method.

With respect to the previous CC-type calculations for boron, highly-accurate non-relativistic calculations  were reported in Ref.~\cite{KloBacTew10}.
These authors included excitations of all five electrons (i.e., CC with single, double, triple, quadruple and quintuple excitations)  and also used the Slater-type geminals at the CCSD levels to correct for the incompleteness of their necessarily very limited computational basis. They then added relativistic corrections to their calculations in an \emph{ad hoc} manner. These authors computed the ionization potential of the ground state of boron and other atoms.
Less complete (CCSD truncation level), albeit fully relativistic calculations were carried out in Ref.~\cite{GhaEliSaf11}. The focus of that work was on the convergence of iteration method for solving linearized CCSD (LCCSD) equations. Here we will use the convergence techniques developed in that work, but will employ a more sophisticated CCSDvT framework.


This paper is organized as follows: In Section~II we discuss the CCSDvT method and justify the use of the convergence method already implemented for LCCSD in Ref.~\cite{GhaEliSaf11}. In Section~III we present calculated energy levels of many of the low-lying states of born, their electric-dipole transition amplitudes and hyperfine constants. We also compare our results with other computational methods and experimental data. Finally, in Section~V we draw the conclusions.

\section{METHOD}
\label{Sec:Method}

\subsection{Coupled-cluster method and approximations}
Here we briefly describe the coupled-cluster method and approximations used in our computations. We use coupled-cluster formalism for systems with one valence electron outside the closed-shell core. The reader may find more detailed descriptions of what follows in Refs.\cite{LinMor86, Lin85, DerEmm02}.\\
In our treatment of the atomic Hamiltonian for one valence electron systems, we employ the frozen-core Dirac-Hartree-Fock (DHF) potential. In the second-quantization notation, the atomic Hamiltonian in the DHF basis, ignoring a common energy offset, reads:
\begin{align}\label{Eq:Hamiltonian}
\widehat{H}=\hat{H}_0+\hat{G}=\sum_i\varepsilon_iN[\hat{a}^\dagger_i \hat{a}_i]+\frac{1}{2}\sum_{ijkl}g_{ijkl}N[\hat{a}^\dagger_i\hat{a}^\dagger_j\hat{a}_l\hat{a}_k]\, .
\end{align}
Here $\hat{H}_0$ is the one-electron lowest order Hamiltonian and $\hat{G}$ is the residual Coulomb interaction. $\varepsilon_i$ is the single-particle DHF energy and $g_{ijkl}$ is the two-body Coulomb matrix element. $N[...]$ indicates that the operators are in normal form with respect to the quasi-vacuum core state, $|0_c\rangle$. Operators $\hat{a}^\dagger$ and $\hat{a}$ are respectively creation and annihilation operators.\\
In our implementation of the coupled-cluster method the exact atomic wave function of a monovalent atom with the valence electron in state $v$ is written as:
\begin{align}\label{Eq:exactwave}
|\Psi_v\rangle=\widehat{\Omega}|\Psi^{(0)}_v\rangle,
\end{align}
 where $|\Psi^{(0)}_v\rangle=\hat{a}_v^\dagger|0_c\rangle$ is the lowest-order DHF wave function. The wave operator $\hat{\Omega}$ maps the DHF solution onto the exact wave function, $|\Psi_v\rangle$. The ansatz for the wave operator is:
\begin{align}\label{Eq:ansatz}
\hat{\Omega}=N[\exp(\hat{C})],
\end{align}
  where $\hat{C}$ is called the cluster operator. For a system of $N$ electrons it is expanded as:
\begin{align}\label{Eq:Opsum}
\hat{C}=\sum_1^N \hat{C}_i\,
\end{align}
  where $i$ indicates the number of excitations of core and valence electrons. For example, the operator $\hat{C}_1$ may be split into two classes of core and valence excitations:
\begin{align}\label{Eq:single}
&\hat{C}_1=\hat{S}_{c}+\hat{S}_{v}, \\ \nonumber
&\hat{S}_{c}=\sum_{ma}\rho_{ma}\hat{a}^\dagger_m\hat{a}_a, \\ \nonumber
&\hat{S}_{v}=\sum_{mv}\rho_{mv}\hat{a}^\dagger_m\hat{a}_v\, .
\end{align}
The coefficients $\rho_{ma}$ and $\rho_{mv}$ above are referred to as cluster amplitudes and are to be found. Here and elsewhere in this paper, the indices $a, b, ...$ are reserved for core orbitals, $m, n, ...$ are designated to virtual or excited orbitals, $v, w, ...$ indicate valence states and $i$, $j$, $k$, and $l$ are arbitrary orbitals. We will subsequently equate $\hat{C}_2$ with $\hat{D}$ and $\hat{C}_3$ with $\hat{T}$. Explicitly,

\begin{align}\label{Eq:double-triple}
&\hat{D}_{c}=\frac{1}{2!}\sum_{mnab}\rho_{mnab}\hat{a}^\dagger_m \hat{a}^\dagger_n \hat{a}_b\hat{a}_a, \\
&\hat{D}_{v}=\sum_{mna}\rho_{mnva}\hat{a}^\dagger_m \hat{a}^\dagger_n \hat{a}_a\hat{a}_v, \\
&\hat{T}_{c}=\frac{1}{3!}\sum_{mnrabc}\rho_{mnab}\hat{a}^\dagger_m \hat{a}^\dagger_n \hat{a}^\dagger_r \hat{a}_c \hat{a}_b \hat{a}_a, \\
&\hat{T}_{v}=\frac{1}{2!}\sum_{mnrab}\rho_{mnrvab}\hat{a}^\dagger_m \hat{a}^\dagger_n \hat{a}^\dagger_r \hat{a}_b \hat{a}_a \hat{a}_v\, .
\end{align}

Expanding the $\exp(\hat{C})$ will lead to various powers and products of different cluster operators. Approximate solutions to the wave function $|\Psi_v\rangle$ are then found by keeping only a certain number of terms.
As an example, the linearized coupled-cluster single-double (LCCSD) method keeps only the $\hat{S}$ and $\hat{D}$ terms:
\begin{align}\label{Eq:LCCSD}
|\Psi_v\rangle=(1+\hat{S}+\hat{D})|\Phi_v\rangle=\hat{\Omega}_{\mathrm{LCCSD}}|\Phi_v\rangle.
\end{align}
Another approximation, coupled-cluster single-double 2 (CCSD2), keeps only up to the second-order terms in $\hat{S}$ and $\hat{D}$ and discards higher orders:
\begin{align}\label{Eq:CCSD2}
|\Psi_v\rangle=(1+\hat{S}+\hat{D}+\frac{1}{2}\hat{S}^2+\frac{1}{2}\hat{D}^2+\hat{S}\hat{D})|\Phi_v\rangle=\hat{\Omega}_{\mathrm{CCSD2}}|\Phi_v\rangle.
\end{align}
 We used a combination of the above approximations with the addition of certain contributions of the triple term, $\hat{T}$. Below we will discuss the approximate method used in more detail.

In order to find the cluster-amplitudes, $\rho$'s, one uses a set of generalized Bloch equations, see Ref.~\cite{DerEmm02} for equations specific to a monovalent system. There are two such sets of equations for the core and valence states. The first set involves only the core-related operators. This set can be written as:
\begin{align}\label{Eq:coreBloch}
(\varepsilon_v-\hat{H}_0)\hat{\Omega}^{\mathrm{core}}|\Psi^{(0)}_v\rangle=(\{\hat{Q}\hat{G}\hat{\Omega}\}_{\mathrm{conn.}})^{\mathrm{core}}|\Psi_v^{(0)}\rangle\,.
\end{align}
 Here $\hat{Q}=1-|\Psi^{(0)}_v\rangle\langle\Psi^{(0)}_v|$ is a projection operator and $\varepsilon_v$ is the Hartree-Fock energy of the valance electron. The subscript ``conn.'' indicates that the retained Brueckner-Goldstone diagrams have no disconnected parts except for the valence lines. The above core equation does not depend on the valence state.

The other set of Bloch equations is formulated for valence amplitudes,
\begin{align}\label{Eq:valBloch}
(\epsilon_v+\delta E_v-\hat{H}_0)\hat{\Omega}^{\mathrm{val}}|\Psi^{(0)}_v\rangle=(\{\hat{Q}\hat{G}\hat{\Omega}\}_{\mathrm{conn.}})^{\mathrm{val}}|\Psi_v^{(0)}\rangle\,.
\end{align}
 Here $\delta E_v=\langle\Psi^{(0)}_v|\hat{G}\hat{\Omega}|\Psi^{(0)}_v\rangle$ is the correlation energy. It should be noted that the right-hand side of the Eq.~(\ref{Eq:valBloch}) contains both core and valence amplitudes.

 In order to solve the Bloch equations (\ref{Eq:coreBloch}) and (\ref{Eq:valBloch}), one could devise an iterative approach. Doing this we obtain the recursive relation:
\begin{align}\label{Eq:recBloch}
\left(\epsilon_v+(\delta E_v)-\hat{H}_0\right)\hat{\Omega}^{(n+1)}|\Psi^{(0)}_v\rangle=(\{\hat{Q}\hat{G}\hat{\Omega}^{(n)}\}_{\mathrm{conn.}})|\Psi_v^{(0)}\rangle,
\end{align}
 with $\hat{\Omega}^{(0)}=1$. In the above equation $(\delta E_v)$ means that the $\delta E_v$ should only be kept in the valence equation, (\ref{Eq:valBloch}). Equations (\ref{Eq:coreBloch}) to (\ref{Eq:recBloch}) are general and one can substitute different states in them.

  The core equations are solved by using the CCSD approximation, Eq.~(\ref{Eq:CCSD2}). In the present work, we used the nonlinear terms in Eq.~(\ref{Eq:CCSD2}) in addition to some leading terms of the valence triples of Eq.~\ref{Eq:valT}. All single and double contributions of the CCSD method (including the CCSD2 terms) are spelled out in detail in Ref.~\cite{PorDer06Na} and the triple terms, including some we have discarded, are shown graphically in Ref.~\cite{DerPorBel08}. Therefore, here we only discuss such contributions qualitatively.

   Both core and valence Bloch equations can be further separated into equations for single, double, and triple cluster amplitudes. For example, the topological structure of valence singles equation becomes~\cite{DerPorBel08}:
\begin{align}\label{Eq:valS}
-[\hat{H}_0,\hat{S}_{v}]+\delta E_v \hat{S}_{v}=\mathrm{CCSD2}+\hat{S}_{v}[\hat{T}_{v}],
\end{align}
 where the notation $\hat{S}_{v}[\hat{T}_{v}]$ stands for the effect of valence triples ($\hat{T}_{v}$) on the right-hand side of valence singles ($\hat{S}_{v}$) equation. It should be noted that in the equation above and what follows $\delta E_v=\delta E_{\mathrm{CCSD2}}+\delta E_v[\hat{T}_{v}]$ is the correlation energy. The equation for valence doubles reads:
\begin{align}\label{Eq:valD}
-[\hat{H}_0,\hat{D}_{v}]+\delta E_v \hat{D}_{v}\approx \mathrm{CCSD2}+\hat{D}_{v}[\hat{T}_{v}]\,.
\end{align}
 Finally the approximation used for valence triple equation reads:
 \begin{align}\label{Eq:valT}
-[\hat{H}_0,\hat{T}_{v}]+\delta E_v \hat{T}_{v}\approx \hat{T}_{v}[\hat{D}_{c}]+\hat{T}_{v}[\hat{D}_{v}]+\hat{T}_{v}[\hat{T}_{v}]\,.
\end{align}
In the present work we use two different approximations for the right-hand-side of  the triple equations, Eq(\ref{Eq:valT}). If we keep only the two first terms on the right-hand-side of Eq.~(\ref{Eq:valT}), we call the method CCSDvT2, and if we additionally include the $\hat{T}_{v}[\hat{T}_{v}]$ terms, the method is called CCSDvT3. The reason for such a distinction is that the inclusion of $\hat{T}_{v}[\hat{T}_{v}]$ terms on the right-hand-side of Eq.~(\ref{Eq:valT}), greatly increases the computational demands.

  We further simplify  Eqs.(\ref{Eq:valS})-(\ref{Eq:valT}) by taking advantage of the spherical symmetry of atoms and analytically summing over the magnetic quantum numbers. This results in a reduced form of cluster amplitudes and Coulomb integrals. We use a different notation for such reduced amplitudes. For example, the reduced valence amplitudes of singles are denoted as $\rho(mv)$, doubles as $\rho_k(mnva)$ and triples as $\rho_{k_1k_2h}(mnrvab)$. Here $k_1$, $k_2$ are integer coupling momentum numbers and $h$ is a half integer coupling angular momentum. As an example, the relationship between the ordinary and the reduced triples maybe represented as~\cite{PorDer06Na}

   \begin{equation}
 \rho_{mnrvab}=\sum_{k_1k_2h}%
\raisebox{-6ex}{\includegraphics[
scale=0.5
]{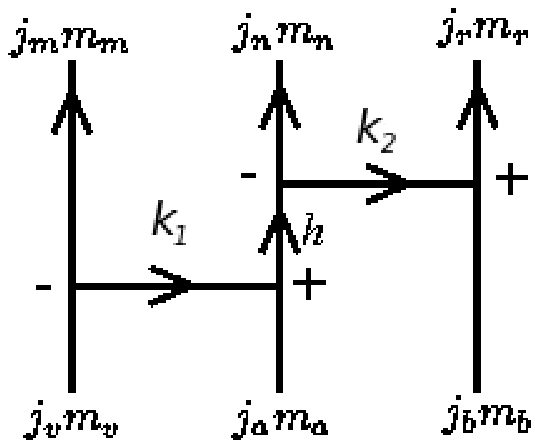}}
\rho_{k_1k_2h}\left(  mnr \, vab\right) \,,
\label{Eq:vTAngular}
\end{equation}
where the diagram subsumes various 3j symbols~\cite{LinMor86}.
Interested readers can find complete discussions of angular reduction in Refs.~\cite{LinMor86} and \cite{Saf00}, for single and double term reductions, and Ref.~\cite{PorDer06Na} for triple term reductions.


%

The angularly reduced equations have numerous terms on their right-hand sides and are composed of the summation of products of cluster amplitudes, $\rho$'s with each other and with two-body Coulomb matrix elements ($g$'s in Eq.~(\ref{Eq:Hamiltonian}).) The right-hand side terms can be found in Refs.~\cite{BluJohLiu89,SafDerJoh98,PorDer06Na} and we will not be reproducing them here. After the angular reduction, the three valence Bloch equations (\ref{Eq:valS})-(\ref{Eq:valT}) can be represented in a general form as:
\begin{widetext}
\begin{subequations}
\begin{align}\label{Eq:clusters}
 &(\varepsilon_m-\varepsilon_v+\delta E_v)\rho(mv)=\sum_{n} A_S \cdot\rho(nv) +\sum_{nak} B_S \cdot \tilde{\rho}_k(mnva) +\sum_{nrabkk'h} C_S \cdot \tilde{\rho}_{k_1k_2h}(mnrvab)+\sum D_S, \\
 &(\varepsilon_{mn}-\varepsilon_{va}+ \delta E_v)\tilde{\rho}_k(mnva)=\sum_{n} A_D \cdot\rho(nv) +\sum_{nak} B_D \cdot \tilde{\rho}_k(mnva) +\sum_{nrabkk'h} C_D \cdot \tilde{\rho}_{k_1k_2h}(mnrvab)+\sum D_D, \\
 &(\varepsilon_{mnr}-\varepsilon_{vab}+ \delta E_v)\tilde{\rho}_{k_1k_2h}(mnrvab)=\sum_{nak} B_T \cdot \tilde{\rho}_k(mnva) +\sum_{nrabkk'h} C_T \cdot \tilde{\rho}_{k_1k_2h}(mnrvab)+\sum D_T,
\end{align}
\end{subequations}
\end{widetext}
where the amplitudes with tilde signs ($\tilde{\rho}$) are antisymmetrized combinations of the reduced amplitudes.
In the above equations $\varepsilon_{ijk}=\varepsilon_i+\varepsilon_j+\varepsilon_k$ and the $A$, $B$, $C$, and $D$'s are constants with their subscripts denoting to which set of equations they belong, for example $S$ for singles etc. It must be emphasized that the right-hand sides of the above equations are linear in terms of valence cluster amplitudes and the constants include already known core cluster amplitudes and Coulomb matrix elements, $g$'s. On the left-hand side, however, the terms are not linear in terms of valence cluster amplitudes due to the presence of $\delta E_v$ terms, which also depend on the valence cluster amplitudes. Despite this fact, in each iteration the $\delta E_v$'s are taken from the previous iteration and substituted to find a new set of cluster amplitudes and $\delta E_v$'s. In this sense the above equations are overall treated as if they were linear at each iteration.

\subsection{Convergence}
Simply iterating the CC recursive equation,  Eq.(\ref{Eq:recBloch}), at times proves to be insufficient as they could lead to slow convergence, non-convergence, or even convergence to nonphysical solutions. In Ref.~\cite{GhaEliSaf11}, we discussed two convergence methods developed in quantum chemistry \cite{PurBar81} and their effectiveness in application to the LCCSD method. Here we apply one of the methods, the direct inversion of iterative space (DIIS) to our CCSDvT2 and CCSDvT3 methods.

To see how the DIIS is applied, we rewrite Eq.(\ref{Eq:clusters}) in a more streamlined fashion:
\begin{align}\
\notag
 \textbf{t}=&\left(\begin{array}{c}
                      \rho(mv) \\
                     \tilde{\rho}_k(mnvb) \\
                      \tilde{\rho}_{k_1k_2h}(mnrvab)\\
                    \end{array}
                  \right)\,
\end{align}
and
\begin{align}\notag
\textbf{a}=&\left(\begin{array}{c}
                      \sum D_S \\
                      \sum D_D \\
                      \sum D_T
                    \end{array}
                  \right), ~~~
                   \textbf{D}=\left(\begin{array}{c}
                      \varepsilon_v-\varepsilon_m + \delta E_v\\
                      \varepsilon_{vb}-\varepsilon_{mn} + \delta E_v\\
                      \varepsilon_{vba}-\varepsilon_{mnr} + \delta E_v\\
                    \end{array}
                  \right).
\end{align}
Then the combined three sets of Eq.(\ref{Eq:clusters}) can be written as:
\begin{align}\label{Eq:combinedval}
\textbf{D}\cdot \textbf{t}=\textbf{a}+\Delta \cdot \textbf{t}\,,
\end{align}
where $\Delta$ is a matrix including all the coefficients in front of cluster amplitudes on the right-hand side of Eq.~(\ref{Eq:clusters}). The above equation can be solved iteratively by re-writing it as
\begin{align}\label{Eq:ite}
\textbf{t}^{(m+1)}=\textbf{D}^{-1}(\textbf{a}+\Delta \cdot \textbf{t}^{(m)})\,.
\end{align}
The iterative equation above is initiated by letting $t^{(0)}=0$ on the right-hand side and finding $t^{(1)}$ and so on.

The DIIS method works in two steps. First, a few iterative solutions $t^{(i)}$ of Eq.(\ref{Eq:ite}) are found. Next, a linear combination of the said $t^{(i)}$ is used to find the best next solution to the equation. For example, after accumulating $m$ iteratively found solutions, $t^{(1)}$, $t^{(2)}$,..., $t^{(m)}$, the next best approximation can be found as their linear combination,
\begin{align}\label{Eq:linearcomb}
\textbf{t}^{(m+1)}=\sum _{i=1}^{m} \sigma_i \textbf{t}^{(i)}=\sigma \cdot \textbf{T}\,.
\end{align}
Here $\sigma_i$ is the weight assigned to $t^{(i)}$. In the case of the DIIS method, the $\sigma$ coefficients are determined by applying an error minimization scheme and solving the resulting system of linear equations \cite{GhaEliSaf11}
\begin{align}\label{Eq:diis}
\textbf{T}^T(\Delta-\textbf{D})^T\textbf{a}+\textbf{T}^T(\Delta-\textbf{D})^T(\Delta-\textbf{D})\textbf{T}\sigma=0\,.
\end{align}
  The new answer thus found, $t^{(m+1)}$ is then fed back to the right-hand side of Eq.(\ref{Eq:ite}) and the two steps are repeated until some parameter, i.e. the $\delta E_v$, stops changing (up to a specified accuracy) between two consecutive iterations.

\subsection{Matrix elements}
After finding the cluster amplitudes and correlation energies, we can calculate matrix elements of a one-particle operator
\begin{align}\label{Eq:MatrixOp}
\hat{Z}=\sum_{ij}z_{ij}\hat{a}_i^\dagger \hat{a}_j,
\end{align}
 where $z_{ij}$ is the single-particle matrix element. Notice that in deriving the CC equations one uses the intermediate normalization scheme $\langle\Psi_v|\Psi^{(0)}_v\rangle=1$. The matrix elements then have the form,
\begin{align}\label{Eq:MatrixEl}
M_{wv}=\frac{\langle\Psi_w|\hat{Z}|\Psi_v\rangle}{\sqrt{\langle\Psi_w|\Psi_w\rangle\langle\Psi_v|\Psi_v\rangle}}\,.
\end{align}
As discussed in Ref.~\cite{BluJohLiu89}, this matrix element could be separated into two parts leading to the expression
\begin{align}\label{Eq:MatrixElSep}
M_{wv}=\delta_{wv}(Z_0)_{\mathrm{conn}}+\frac{(Z_{1})_{\mathrm{conn}}}{\sqrt{[1+(\delta N_w)_{\mathrm{conn}}][1+(\delta N_v)_{\mathrm{conn}}]}}\,.
\end{align}
Here $Z_{0}=\langle0_c|\hat{\Omega}^\dagger \hat{Z} \hat{\Omega}|0_c\rangle$ and the remaining contributions of $Z_{wv}=\langle0_c|\hat{a}_w\hat{\Omega}_{w}^\dagger \hat{Z} \hat{\Omega}_{v}\hat{a}^\dagger_v|0_c\rangle$ are encapsulated into $Z_{1}$. In a similar way
\begin{align*}
 N_w&=\langle\Psi_w|\Psi_w\rangle=N_0+\delta N_w,\\ \nonumber N_0&=\langle0_c|\hat{\Omega}^\dagger \hat{\Omega}|0_c\rangle=1,
  \end{align*}
  with $\delta N_w$ containing the rest of the contributions to the normalization. $Z_0$ contributions vanish for nonscalar  operators $\hat{Z}$, so we can ignore them here.
  In Refs.~\cite{BluJohLiu89,BluJohSap91} the contributions to matrix elements in the LCCSD approximation are explicitly listed. Ref.~\cite{DerPor05} further discusses all the leading nonlinear contributions and the contributions of connected triple excitations to matrix elements.

  When the CC exponent of Eq.(\ref{Eq:ansatz}) is expanded in Eq.(\ref{Eq:MatrixEl}), an infinite number of terms are produced. The resulting series may be partially summed so that it subsumes an infinite number of terms. This procedure is called ``dressing" and is explicitly explained in Ref.~\cite{DerPor05}. In short, it is built on the expansion of the products of the CC-amplitudes into a sum of n-body insertions. Two types of insertions are considered: particle (hole) line insertion and two-particle (two-hole) random-phase-approximation-like insertion. It must be noted that this procedure is specialized for the CCSD approximation in monovalent systems.

  Due to the approximate nature of the CCSDvT method, certain correlation effects are lost. To partially account for the missing contributions in the calculation of the matrix elements, we correct the wave functions using a semi-empirical procedure suggested in Ref.~\cite{BluJohSap92}. In this procedure the valence singles are re-scaled by the ratio of experimental and theoretical correlation energies. In the present paper we refer to the results obtained by such a procedure as ``scaling".

 In the next section we will discuss the results obtained with different approximations and present valence energies, matrix elements, and hyperfine constants for the boron atom. We will also compare our results with previous experimental and theoretical results.

\section{RESULTS AND DISCUSSION}
\label{Sec:Results}
So far we recapitulated various CC approximations. We also described how we apply the DIIS converging method to the iterative solutions of  the LCCSD, CCSD and  CCSDvT equations. In this section we present  \emph{ab initio} numerical calculations for properties of several low-lying levels of the boron atom.

Atomic boron has three valence electrons, with the ground state configuration  $1s^2 2s^2 2p^1$.
In our calculations, we start by assigning the two electron $2s^2$ valence orbitals of the ground state to the core orbitals. Therefore, we approximate the boron atom as a monovalent system.

We employ the dual-kinetic balance B-spline basis set, obtained by solving the frozen core DHF equations \cite{BelDer08}. This basis set numerically approximates a complete basis for the single-particle atomic states. The basis set was generated in a cavity of radius 40 a.u. and contains 40 orbitals per partial wave for energies above the Dirac sea.

The CC core equations are solved in the CCSD approximation.
The core amplitudes are computed with partial waves summed up to and including the angular momentum $l_{max}=6$.
The single and double core excitation coefficients are then fed into the valence equations.
The valence equations are initially solved in the LCCSD approximation.
The resulting LCCSD wave functions are then used to initiate the nonlinear CCSD method.
The valence CCSD result, in turn, becomes the reference point for the CCSDvT computations.
In calculating the valence wave function, basis functions with $l_{max}=6$ are still used for singles and doubles.
However, due to the computational expense of the triple terms, we employed a more limited basis set with $l_{max}=3$ for the triples. Initial calculations with $l_{max}=4$ for all triple terms showed very little change in energies at a great expense in time. Therefore, to keep consistency of the data, we only present results with $l_{max}=3$ for the triple terms.
Also, to keep the computational cost lower, we employed 35 out of 40 positive energy basis functions for the single and double terms, while reduced this number to 25 for the triples. Again, calculations of triples with 35 basis functions showed very little change in the outcome. We further introduced basis extrapolation corrections, $l_{max} \rightarrow \infty$, which is discussed in subsection A.

 In the present work we use two different approximations for triple equations.
 If we only keep the $\hat{T}_v[\hat{D}_c]$ and $\hat{T}_v[\hat{D}_v]$ terms on the right-hand-side of the Eq.~(\ref{Eq:valT}), we call the method CCSDvT2. On the other hand, if we keep all the terms in the Eq.~(\ref{Eq:valT}) we call the method CCSDvT3 with $\hat{T}_{v}[\hat{T}_{v}]$ being the difference with CCSDvT2.
 The inclusion of $\hat{T}_{v}[\hat{T}_{v}]$ terms on the right-hand-side of Eq.~(\ref{Eq:valT}), greatly increase the computational cost of triple calculations.

A brief analysis of the largest valence amplitudes ($\rho's$,) will follow in the next section (Sec.~\ref{SubSec:Terms}).
Results for removal energies are presented in Sec.~\ref{SubSec:energies} and for dipole matrix elements and magnetic-dipole hyperfine constants in Sec.~\ref{SubSec:EDMandHFC}.

\subsection{Largest contributions by cluster amplitude}
\label{SubSec:Terms}
 \begin{figure}
\begin{center}
\includegraphics[scale=0.7]{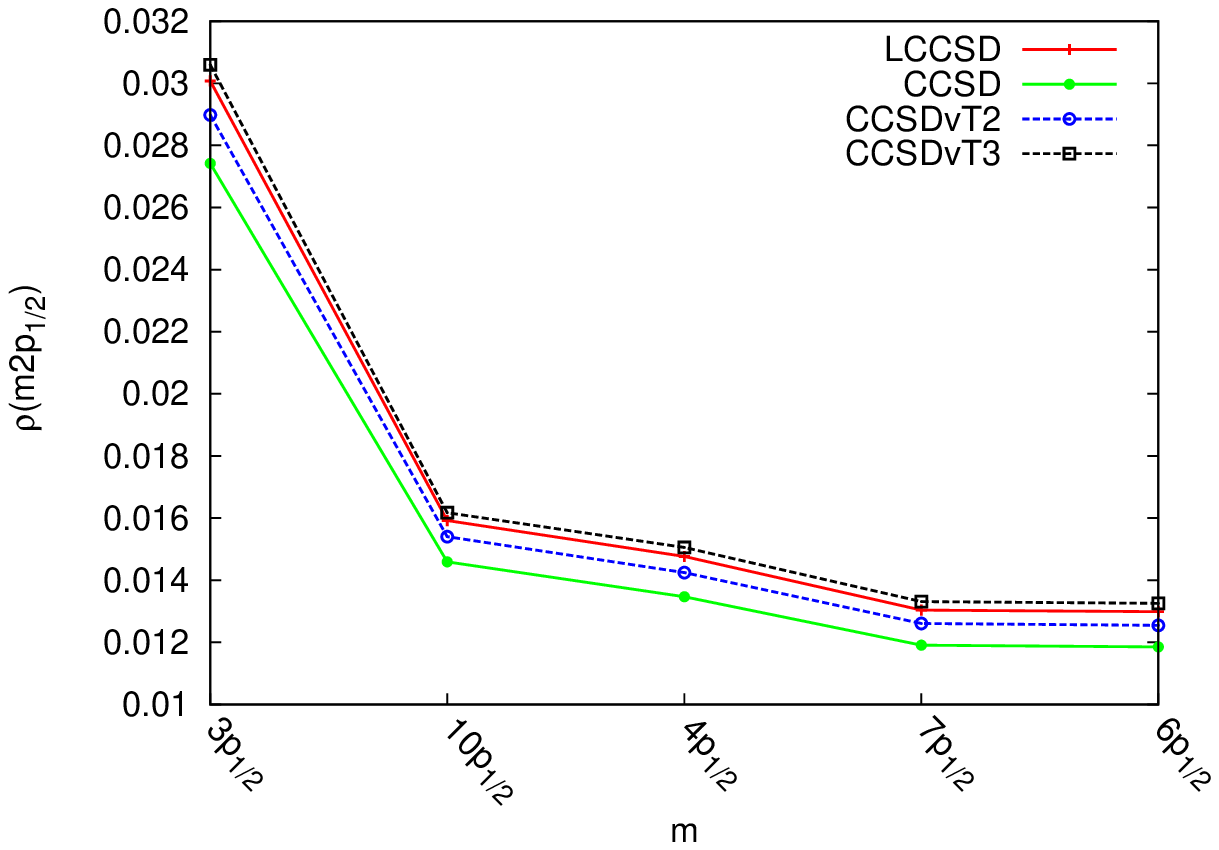}
\end{center}
\caption[]{The five largest reduced single excitation coefficients $\rho(mv)$ are compared in different CC approximations. Here $v$ is the valence orbital $2p_{1/2}$ and $m$ is the excited orbital indicated on the x-axis of the graph.}\label{Fig:singles}
\includegraphics[scale=0.7]{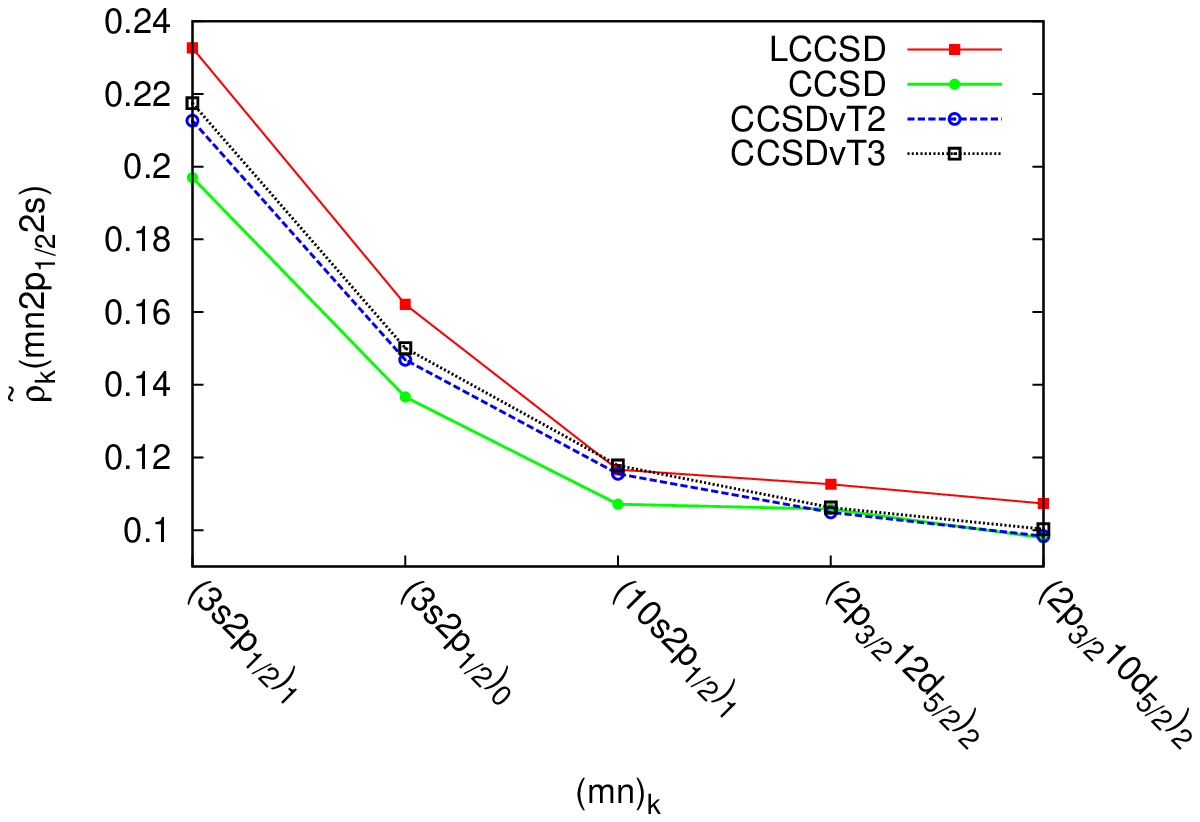}
\caption[]{The five largest anti-symmetrized reduced double amplitudes, $\tilde{\rho}_k(mnva)$, of the ground state of boron are compared in different CC approximations. Here $v$ is the valence orbital $2p_{1/2}$ and $a$ is the core orbital $2s$. Excited orbitals $m$ and $n$ and the angular momentum value $k$ of the reduced amplitudes are shown on the x-axis.}\label{Fig:doubles}
\end{figure}
 \begin{figure}
\begin{center}
\includegraphics[scale=0.7]{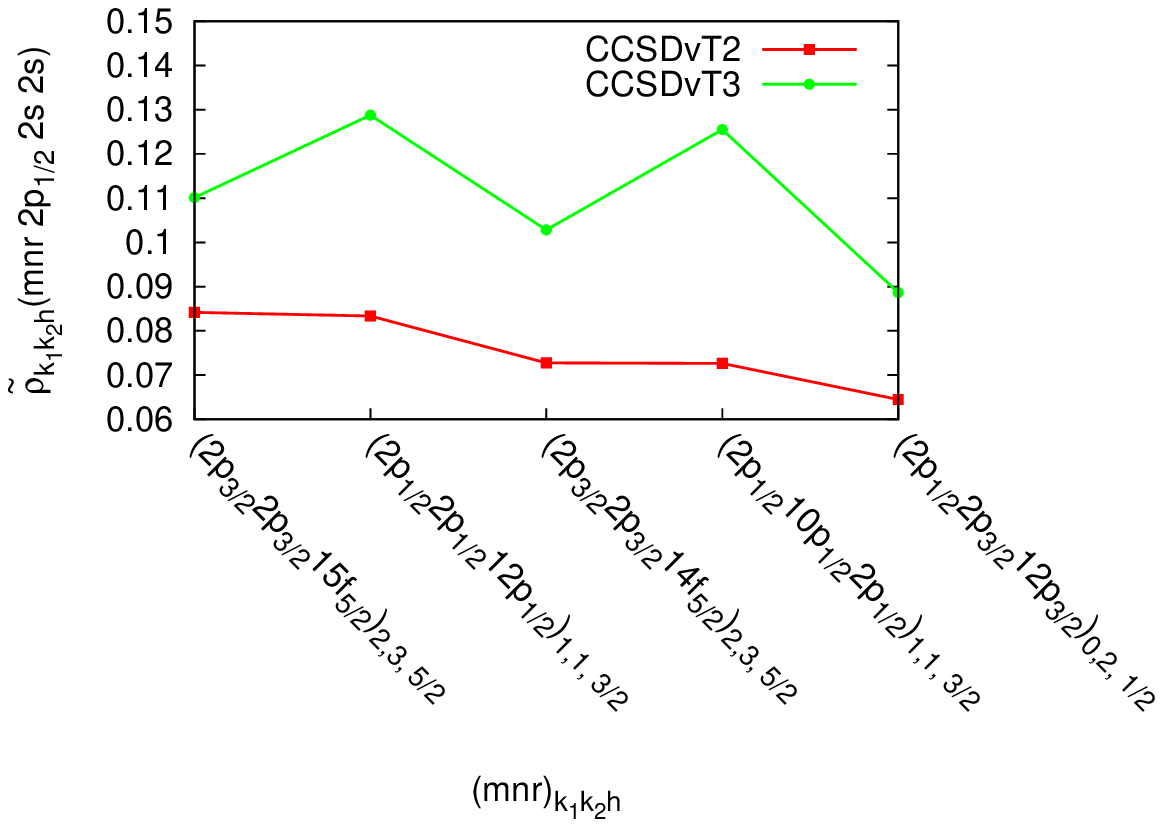}
\end{center}
\caption[]{The five largest reduced triple amplitudes, $\rho_{k_1k_2 h}(mnrvab)$, for the ground state of boron are compared in different CC approximations. Here $v$ is valence orbital $2p_{1/2}$ and $a$ and $b$ are the core orbital $2s$. Excited orbitals $m$, $n$, and $r$ and the coupling angular momentum values $k_1$, $k_2$ plus the half integer $h$ of the reduced triple amplitudes are shown on the x-axis.}
\label{Fig:triples}
\includegraphics[scale=0.7]{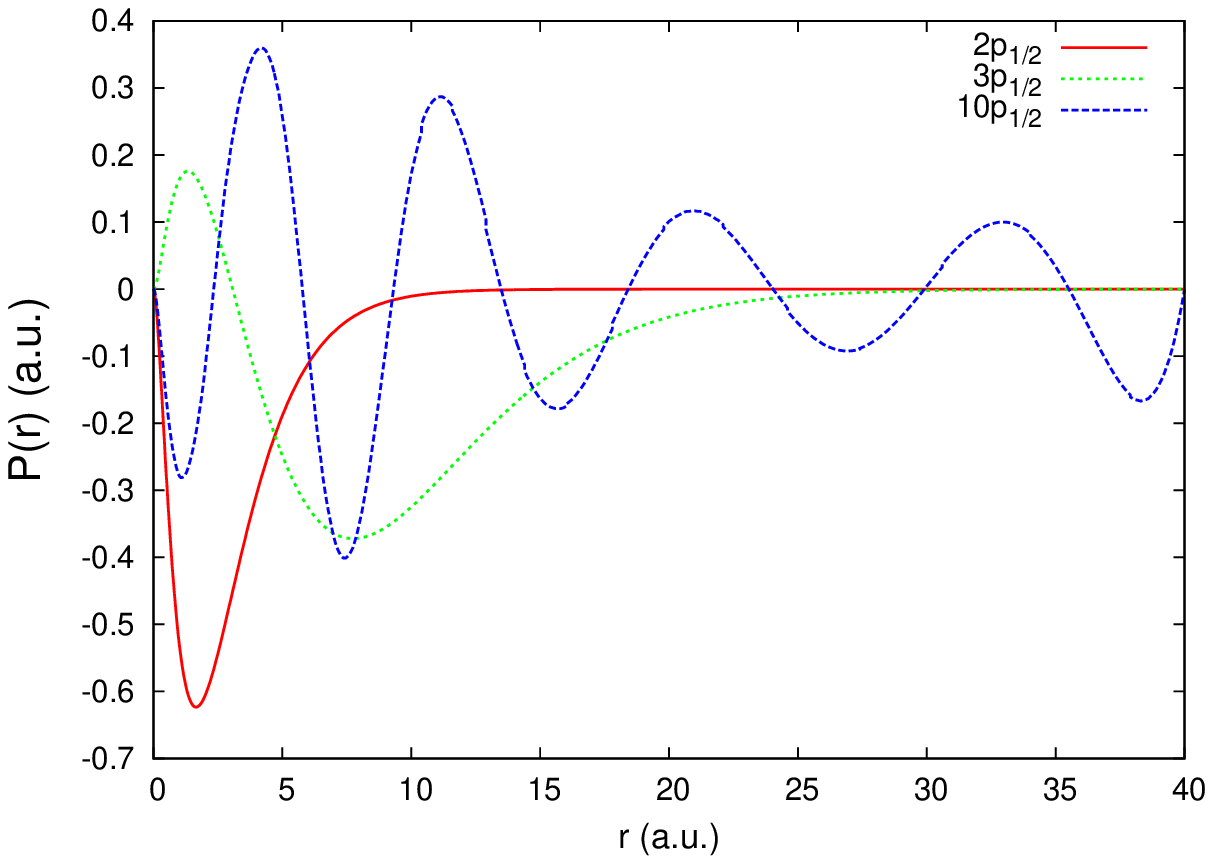}
\caption[]{Large component, $P$, of the orbitals $2p_{1/2}$, $3p_{1/2}$, and $10p_{1/2}$ for boron is shown as a function of the distance from nucleus. The $10p_{1/2}$ wave function is squeezed by the cavity wall (40 a.u. here.) This makes the value of the $\rho(10p_{1/2}, 2p_{1/2})$ amplitude comparable with $\rho(3p_{1/2}, 2p_{1/2})$ (Fig.~\ref{Fig:singles}).}
\label{Fig:splines}
\end{figure}

Before we present the results of the numerical calculations, we would like to investigate the relative importance of individual valence amplitudes ($\rho$'s) for each level of approximation.
We run the numerical code, extract the five largest reduced valence amplitudes, and analyze how they change with each CC approximation. The results are shown in Figs. (\ref{Fig:singles}-\ref{Fig:triples}).
Here, we only analyze the valence single, double and triple amplitudes for the ground state (2$p_{1/2}$) of boron.

 The valence states with the largest amplitudes remain for the most part the same from one approximation to another.  For example, the CCSD approximation will render the same five largest $\tilde{\rho}_k(mnva)$'s as the CCSDvT2 approximation and so on. As is seen in Figs.(\ref{Fig:singles}-\ref{Fig:doubles}), in the CCSD approximation, the largest reduced valence amplitudes are lowered in value as compared to their LCCSD counterparts. Including triples in the CCSDvT2 and CCSDvT3 approximations brings up these values to the region between the LCCSD and the CCSD approximations.

Based on perturbation theory, the cluster amplitudes are in general proportional to the ratio of Coulomb integrals (or their products) to energy differences between the orbitals. Therefore, we expect the orbitals with energies closer to the valence state under study to have the largest contributions to the valence amplitudes. However, as is evident from the figures, some of the larger contributions belong to states with high principal quantum numbers. For example, in Fig.~\ref{Fig:singles}, the $\rho(10p_{1/2}, 2p_{1/2})$ amplitude is the next largest after the $\rho(3p_{1/2}, 2p_{1/2})$. In order to understand the unusually large contributions from highly excited orbitals, we plot the large component, $P(r)$, of the $2p_{1/2}$, $2p_{3/2}$ and $10p_{1/2}$ orbitals in Fig.(\ref{Fig:splines}), where $r$ is the distance from nucleus. Examination of our DHF energies indicates that the continuum states start from the $5p_{1/2}$ orbital. The cavity size and its boundary conditions have the effect of compressing the continuum's $10p_{1/2}$ wave function wave function toward the nucleus. In effect the $10p_{1/2}$ mimics the behavior of $2p_{1/2}$ in lower $r$ regions, which in turn results in large Coulomb integrals between these states. The large contributions of other orbitals with high principal quantum numbers to the amplitudes is the result of such a cavity squeeze on the continuum wave functions. If we were to choose a different cavity radius, some other continuum states would see a similar effect.
%
%



\subsection{Energies}
\label{SubSec:energies}
 The computed energies of the 2$p_{1/2}$, 2$p_{3/2}$, 3$s_{1/2}$, 3$p_{1/2}$, 3$p_{3/2}$ and 4$s_{1/2}$ valence states of neutral boron are compiled in Table~\ref{Table1}. We compare our results with the National Institute of Standards and Technologies (NIST) recommended values~\cite{NIST_ASD}. In Table~\ref{Table1}, we give a breakdown of the contribution to the valence energies at each level of the CC approximation. For example, the energy difference between the DHF and the LCCSD methods is shown in the +LCCSD column, while the energy difference between the LCCSD and the CCSD approximations is written under the +CCSD column heading and so on. The extrapolation values are calculated using the results in the LCCSD approximation. We first run the LCCSD code, limiting the angular momenta from $l_{max}$=1 to 6. Next we extrapolate the correlation energy difference between each successive pairs of angular momentum numbers by using the model discussed in Ref.~\cite{BluJohLiuSap89}. The values labeled ``Total" are equivalent to the CCSDvT3 values or the addition of the contributions from all of the previous columns.
 Most of the correlation energy is recovered in the LCCSD method. Generally, as the complexity of the method grows, the respective correlation energy contributions become smaller. However,
as can be seen from Table~\ref{Table1}, the energy contributions of the CCSDvT2 and CCSDvT3 methods are comparable to each other. The percentage difference between our final CCSDvT3 results and the NIST values are between 0.2\% and 0.4\%.

  In Table~\ref{Table2}, the same results are shown again, but this time the valence energies are tabulated with respect to the 2$p_{1/2}$ ground state.
 The results in Table~\ref{Table2} are compared with the NIST recommended values, as well as two other theoretical calculations.
 The Gaussian-correlated (GC) method of Ref.~\cite{BubAda11} is a non-relativistic approach, therefore, it does not resolve the fine-structure splitting for the $p$ states. Multiconfiguration Hartree-Fock (MCHF) results of Ref.~\cite{FroTac04} are calculated relativistically by using the Breit-Pauli Hamiltonian. The MCHF result for the 4$s_{1/2}$ state, however, was not included in Ref.~\cite{FroTac04}, therefore, an older MCHF result~\cite{CarJonStu94} by the same group was used for comparison.

  The most accurate coupled-cluster calculation for the ground state of boron to date has been carried out by Klopper \emph{et al.}~\cite{KloBacTew10}. The starting point of their method is known in the literature as the CCSD(F12) method. In their computational approach, the single-particle basis sets are substantially truncated and are in effect incomplete. The incompleteness of the basis set is compensated by accounting for additional double excitations into Slater-type geminals (which is indicated by the F12 qualifier) at the CCSD level of approximation. At this level of approximation our two methods should be equivalent to each other, since we saturate our basis sets by carrying out extrapolations to higher partial waves. However, comparison of our and Ref.~\cite{KloBacTew10} results at individual levels of approximations is meaningless, since the starting point of the CC method, the independent-particle approximation, in our respective approaches is different. To obtain their high accuracy results, Ref.~\cite{KloBacTew10} include CC up to 5-fold connected excitations. In addition, they add relativistic corrections to their non-relativistic calculations results in an \emph{ad hoc} manner.
    Their final result for boron's $2p_{1/2}$ state is -66934.4 $\mathrm{cm}^{-1}$.
 Nevertheless our CSDvT result without extrapolations is about 100 $\mathrm{cm}^{-1}$ and with extrapolation 40 $\mathrm{cm}^{-1}$ off the experimental value, while the final results of Ref.~\cite{KloBacTew10} recovers the experimental value. This indicates the important role  of omitted higher-order terms (quadrupole and quintuple excitations).

\begin{table*}[ht]%
\addtolength{\tabcolsep}{8.5pt}
\caption{Contributions of each level of approximation to the valence energies of different B valence orbitals are shown and compared with the NIST recommended values~\cite{KraRya07}. All energies are in $\mathrm{cm}^{-1}$ units.}
\begin{ruledtabular}\begin{tabular}{lcccccccc}
B               &$2p_{1/2}$ & $2p_{3/2}$ &$3s_{1/2}$    &$3p_{1/2}$     &$3p_{3/2}$     & $4s_{1/2}$\\
\hline
\\
DHF             &-60546.22  &-60528.30   &-25137.94     &-17258.14      &-17256.30      &-11368.93\\[0.3pc]
+LCCSD          &-6538.07   &-6537.01    &-1966.28      &-1244          &-1243.46       &-155.37\\[0.3pc]
+CCSD           &800.41     &800.14      &378.87        &228.44         &228.34         &-284.29\\[0.3pc]
+CCSDvT2        &-273.05    &-272.39     &-129.89       &-37.24         &-40.14         &-62.92\\[0.3pc]
+CCSDvT3        &-245.35    &-244.97     &-143.8        &-55.3          &-55.5          &-65.38\\[0.3pc]
Extrapolation   &-84.3      &-84.3       &-1.16         &-13.46         &-13.44         &-12.53\\[0.3pc]
Total           &-66886.58  &-66866.83   &-26993.76     &-18379.7       &-18380.5       &-11949.42\\[0.3pc]
NIST~\cite{NIST_ASD}&-66928.04 &-66912.75 &-26888.35    &-18316.17      &-18314.39      &-11971.81\\[0.3pc]
 \end{tabular}\end{ruledtabular}
 \label{Table1}
\end{table*}

%
%

\begin{table*}[ht]%
\addtolength{\tabcolsep}{10pt}
\caption{Comparison of atomic energy levels of B, computed in different coupled cluster approximations, with NIST benchmark values~\cite{KraRya07} and two theoretical results: computational (multiconfiguration Hartree-Fock (MCHF)~\cite{FroTac04} and Gaussian-Correlated (GC)~\cite{BubAda11}). All energies are in $\mathrm{cm}^{-1}$ units. GC method's results are nonrelativistic and do not resolve the fine-structure splitting, therefore there is only one GC value per total angular momentum, $j$.}
\begin{ruledtabular}\begin{tabular}{lccccccc}

  State            & LCCSD  &  CCSD  &  CCSDvT2   &   CCSDvT3  &  MCHF~\cite{FroTac04} &GC~\cite{BubAda11}  &NIST~\cite{NIST_ASD}  \\ \hline \\
$2p_{1/2}$         &    0   &  0     &   0        &    0       &    0                   &   0               &      0                \\
     [0.3pc]
$2p_{3/2}$         & 18.98  &  18.71 &   19.33    &   19.75    &    15.39               &     -             &      15.29            \\
     [0.3pc]
$3s_{1/2}$         &40069.65&39648.11&39791.23    &   39892.82 &    40005.27            &    40048.20       &      40039.69         \\
     [0.3pc]
$3p_{1/2}$         &48652.99&48081.02&48316.79    &   48506.88 &    49011.74            &    48619.04       &      48611.87         \\
     [0.3pc]
$3p_{3/2}$         &48655.39&48083.32&48316.19    &   48506.08 &    49013.49            &     -             &      48613.65         \\
     [0.3pc]
$4s_{1/2}$         &55631.76&54547.06&54757.15    &   54937.16 &    4642.89\footnotemark[1]             &     55017.55      &      55010.23         \\
     [0.3pc]

 \end{tabular}\end{ruledtabular}
  \footnotetext[1]{Ref.~\cite{CarJonStu94}}
  \label{Table2}
\end{table*}
\subsection{Electric-dipole amplitudes and hyperfine constants}
\label{SubSec:EDMandHFC}
After computing the wave functions as described in the previous section, we proceed to calculate the electric-dipole transition amplitudes and the magnetic-dipole hyperfine-structure (HFS) constants. We tabulate the results for the 2$p_{1/2}$, 2$p_{3/2}$, 3$s_{1/2}$, and 4$s_{1/2}$ states of atomic boron. Here we discuss our results and compare them with the MCHF values of Refs.~\cite{FroTac04} and~\cite{CarJonStu94}.

   In Table~\ref{Table3}, we present the reduced electric dipole matrix elements, $\langle3s||\hat{D}||2p_{j}\rangle$ and $\langle4s||\hat{D}||2p_{j}\rangle$, computed in the length form~\cite{Joh07book}. Here, $\hat{D}$ is the electric-dipole operator and $j$=1/2, 3/2. In the columns under the coupled-cluster heading, we list our results in order of the increasing complexity of the employed CC approximations. The corrections to the transition amplitudes, using the dressing and the scaling procedures, are in the next two columns. Our final results for the electric-dipole transition amplitudes are written in the total column. These values are the addition of the corrections to the results obtained in the CCSDvT3 approximation. The MCHF method's result of Refs.~\cite{FroTac04} and~\cite{CarJonStu94} are tabulated in the last column. As can be seen from Table~\ref{Table3}, our final results and the MCHF results are in a reasonable agreement with each other (about 0.3\%.)

  In Ref.~\cite{FroTac04}, the stated values are in the linestrength form, $S_l$, which is the reduced transition amplitude squared. The older Ref.~\cite{CarJonStu94}, contains non-relativistic results for the reduced transition amplitudes between the 4$s$ and the ground state, 2$p$. In the LS coupling scheme, the relativistic and non-relativistic reduced matrix elements, between $s$ and $p$ states, are related as
\begin{subequations}
\begin{align}\label{Eq:RedMatrixConvofDA}
&|\langle n's_{1/2}||\hat{D}||np_{1/2}\rangle|=\sqrt{\frac{2}{3}} |\langle n's||\hat{D}||np\rangle|, \\ \label{Eq:RedMatrixConvofDB}
&|\langle n's_{1/2}||\hat{D}||np_{3/2}\rangle|=\sqrt{\frac{4}{3}} |\langle n's||\hat{D}||np\rangle|.
\end{align}
\end{subequations}
Therefore, we used the above relation to convert the non-relativistic results of Ref.~\cite{CarJonStu94} to relativistic ones.
\begin{table*}[ht!]%
\addtolength{\tabcolsep}{7pt}
\caption{The reduced electric dipole transition-matrix elements $\langle a||\hat{D}||b\rangle$ in the length form in atomic units.}
\begin{ruledtabular}\begin{tabular}{ccccccccccc}

 &&\multicolumn{4}{c}{Coupled Cluster} && \multicolumn{2}{c}{Corrections} &\multicolumn{1}{c}{Total}&\multicolumn{1}{c}{MCHF} \\ \cline{3-6} \cline{8-9} \\

 \multicolumn{1}{c}{$|a\rangle$}&\multicolumn{1}{c}{$|b\rangle$}& \multicolumn{1}{c}{LCCSD} & \multicolumn{1}{c}{CCSD} & \multicolumn{1}{c}{CCSDvT2} & \multicolumn{1}{c}{CCSDvT3}& &$\Delta$(dressing)&$\Delta$(scaling)\\ \hline \\

\multirow{2}{*}{$3s$} &$2p_{1/2}$   &1.2454&1.1772&1.1623&1.1542&&-0.0097&-0.0077&1.1368&1.1345\footnotemark[1]  \\
                      &$2p_{3/2}$   &1.7618&1.6656&1.6444&1.6330&&-0.0143&-0.0105&1.6082&1.6047\footnotemark[1]  \\
     [0.3pc]
\multirow{2}{*}{$4s$}&$2p_{1/2}$ &0.3736&0.4420&0.4498&0.4480&&-0.0027&-0.0067&0.4386&0.4399\footnotemark[2]  \\
                     &$2p_{3/2}$ &0.4791&0.6251&0.6359&0.6331&&-0.0037&-0.0093&0.6201&0.6222\footnotemark[2]  \\
     [0.3pc]

 \end{tabular}\end{ruledtabular}
  \footnotetext[1]{Ref.~\cite{FroTac04}}
\footnotetext[2]{Ref.~\cite{CarJonStu94}}
 \label{Table3}
\end{table*}

A useful test of self-consistency of our results is to check the ratio of the reduced transition amplitudes, $|\langle n's||\hat{D}||np_{1/2}\rangle/\langle n's||\hat{D}||np_{3/2}\rangle|$. This ratio should be equivalent to the ratio of the right-hand-side of Eqs.(\ref{Eq:RedMatrixConvofDA}) and~(\ref{Eq:RedMatrixConvofDB}), or $1/\sqrt{2}\simeq0.7071$.
 The ratio of the first pair of amplitudes in our different approximations is about 0.7068 and for the second pair is 0.7076. These differences can be explained by relativistic corrections.

%
%
%
%
%
%

In Table~\ref{Table4}, we compile the results of calculations of magnetic-dipole hyperfine-structure (HFS) constants, $A$, for the three states 2$p_{1/2}$, 2$p_{3/2}$, and 3$s_{1/2}$ of boron.  Here we list the HFS constants for each level of the CC approximation. The corrections made to each HFS constant by the dressing, scaling, and extrapolation procedures follow the CCSDvT3 approximation's result. Our final results, the MCHF method's $A$ constants~\cite{JonFroGod96} as well as experimental results~\cite{HarEvaLew72} are shown at the bottom of Table~\ref{Table4}. Our 2$p_{1/2}$ hyperfine constant is off by 2\% from the experimental result, while for the 2$p_{3/2}$ state the difference is about 1\%. There are no experimental literature values for the 3$s_{1/2}$  state of boron, so we compare our results with Ref.~\cite{JonFroGod96}'s MCHF value. The percentage difference with MCHF calculation for the 3$s_{1/2}$ orbital's HFS constant is about 0.3\%.

\begin{table}[ht!]%
\addtolength{\tabcolsep}{1pt}
\caption{Changes in valence magnetic-dipole hyperfine-structure constants of 2$p_{1/2}$, 2$p_{3/2}$, and 3$s_{1/2}$ states of boron (in MHz) is shown for different CC approximations.
 A comparison with experimental values of Ref.~\cite{HarEvaLew72})and MCHF method's results  of Ref.~\cite{JonFroGod96} is presented at the bottom bracket.}
\begin{ruledtabular}\begin{tabular}{lccc}
A                       &  2$p_{1/2}$   & 2$p_{3/2}$        &   3$s_{1/2}$ \\
  \hline \\
DHF                     &  317.1        &  63.3             &    146.9      \\
     [0.3pc]
LCCSD                   &  354.3        &  87.0             &    263.2      \\
     [0.3pc]
CCSD                    &  358.6        &  78.0             &    235.3      \\
     [0.3pc]
CCSDvT2                 &  364.1        &  75.0             &    240.2      \\
     [0.3pc]
CCSDvT3                 &  368.0        &  74.8             &    242.8      \\
     [0.3pc]
$\Delta$ (Scaling)      &  0.7          &   0.2             &     -3.2      \\
 [0.3pc]
$\Delta$ (Dressing)     &  4.7          &   -2.3            &     -4.1      \\
 [0.3pc]
$\Delta$(Extrapolation) &  -0.2         &   0.3             &     1.5       \\
 [0.3pc] \hline \\
Final Result            &   373.3       &    72.7            &     235.6     \\
    [0.3pc]
MCHF~\cite{JonFroGod96}  &  366.1        &    73.24           &     234.83   \\
    [0.3pc]
Experimental~\cite{HarEvaLew72}            &  366.0765      &  73.3470       &  -         \\
\end{tabular} \end{ruledtabular}
\label{Table4}
\end{table}
\section{CONCLUSION}
\label{SubSec:Conclusion}
To reiterate, we examined the application of various coupled-cluster approximations for atomic boron. We treated the trivalent boron atom as a monovalent system, taking into account that the 2$p$ valence electron is often the excited electron in optical transitions. We tabulated the results for a few valence energies, electric-dipole transition amplitudes, and magnetic-dipole hyperfine-structure constants of boron in the previous section.
Furthermore, we compared our results with other computational and experimental benchmarks.
The results for the energies were found to be within 0.2\% to 0.4\% of the the NIST recommended values. The results for the electric-dipole transition amplitudes had about 0.3\% difference with the MCHF benchmarks, while our HFS constants differed with the experimental values by 1\% to 2\%.
Considering that 0.1\% accuracies are typical for true monovalent systems (alkali-metal atoms,) the attained 1\% accuracies for boron indicates deficiencies in treating it as a monovalent system.
Indeed the comparison of our results with the more accurate CC computations of Ref.~\cite{KloBacTew10} shows that the way forward may be employing higher rank coupled-cluster amplitudes, quadruples and quintuples.

\section*{Acknowledgments}
This work was supported in part by the US National Science Foundation Grant  No.\ PHY-9-69580.


\end{document}